\documentclass[twocolumn, tighten15]{aastex63}

\usepackage{color}

\received{March 6, 2020}
\revised{April 24, 2020}
\accepted{May 22, 2020}
\submitjournal{ApJL}

\shorttitle{Particle acceleration in kink-unstable jets}
\shortauthors{Davelaar et al.}
\graphicspath{{./}{}}

\begin{document}

\title{Particle acceleration in kink-unstable jets}

\correspondingauthor{Jordy Davelaar}
\email{j.davelaar@astro.ru.nl}

\author[0000-0002-2685-2434]{Jordy Davelaar}
\affiliation{Department of Astrophysics/IMAPP, Radboud University Nijmegen, P.O. Box 9010, 6500 GL Nijmegen, The Netherlands} 
\affiliation{Center for Computational Astrophysics, Flatiron Institute, 162 Fifth Avenue, New York, NY 10010, USA}

\author[0000-0001-7801-0362]{Alexander A. Philippov}
\affiliation{Center for Computational Astrophysics, Flatiron Institute, 162 Fifth Avenue, New York, NY 10010, USA}
\affiliation{Moscow Institute of Physics and Technology, Dolgoprudny, Institutsky per. 9, Moscow region, 141700, Russia}

\author[0000-0003-4271-3941]{Omer Bromberg}
\affiliation{The Raymond and Beverly Sackler School of Physics and Astronomy, Tel Aviv University, Tel Aviv 69978, Israel}

\author[0000-0002-7782-5719]{Chandra B. Singh}
\affiliation{The Raymond and Beverly Sackler School of Physics and Astronomy, Tel Aviv University, Tel Aviv 69978, Israel}



\begin{abstract}

Magnetized jets in GRBs and AGNs are thought to be efficient accelerators of particles, however, the process responsible for the acceleration is still a matter of active debate. In this work, we study the kink-instability in non-rotating force-free jets using first-principle particle-in-cell simulations. We obtain similar overall evolution of the instability as found in MHD simulations. The instability first generates large scale current sheets, which at later times break up into small-scale turbulence. Reconnection in these sheets proceeds in the strong guide field regime, which results in a formation of steep power laws in the particle spectra. Later evolution shows heating of the plasma, which is driven by small-amplitude turbulence induced by the kink instability. These two processes energize particles due to a combination of ideal and non-ideal electric fields.  

\end{abstract}


\section{Introduction}

Magnetized relativistic jets are efficient particle accelerators. They are observed in a broad variety of astronomical sources, e.g., X-ray binaries, Active Galactic Nuclei (AGN), and gamma-ray bursts (GRBs), see e.g. \cite{pudritz2012} for a review on jets. These sources are typically observed over the entire electromagnetic spectrum from radio to $\gamma$-rays, and are considered as main candidates for accelerating ultra-high-energy cosmic rays. Their observed spectral energy distributions suggest that a large fraction of the radiatively important electrons are non-thermal. However, the way these jets accelerate electrons is still uncertain. An effective mechanism for particle acceleration in highly magnetized flows is the dissipation of magnetic energy via reconnection in thin current sheets \citep{zenitani2001,cerutti2014,sironi2014,guo2014}. The reconnection is driven by the plasmoid instability \citep{loureiro2007}, which continuously breaks current sheets into plasmoids separated by X-points. In the case of relativistic reconnection, strong electric fields in the vicinity of X-points accelerate electrons up to $\gamma_{\rm max} \approx 4\sigma$ \citep{werner2016}, where $\sigma={B^2}/({4\pi m_{\rm e} n}c^2)$, $B$ is the magnetic field strength, $m_{\rm e}$ is the electron mass, and $n$ is the plasma number density. A secondary acceleration phase that happens inside the plasmoids pushes particles to higher energies \citep{petropoulou2018}. The study of  reconnection is usually done with kinetic plasma simulations, which model reconnection from first principles by using Harris sheets as initial conditions. However, it is still unknown if and where such sheets can form in realistic jets, and what the geometry of the reconnecting magnetic field is.

Global magnetohydrodynamics (MHD) simulations show that near the launching site jets expand and quickly loose transverse causal contact, making them stable for current-driven instabilities \citep{tchekhovskoy2016,bromberg2016}. As the pressure of the confining medium becomes important, the flow is re-collimated and regains causal contact. As a result, the toroidal hoop stress becomes effective, and compresses the flow into forming a nozzle, which may become prone to internal kink-instability. In the context of astrophysical jets, the kink instability is generally divided into two types: {\it internal kink}, which grows at the jet's core and is not affecting the jet boundaries and, {\it external kink}, which grows on the jet boundaries and perturbs the entire jet body. \cite{bromberg2016} and \cite{tchekhovskoy2016} showed that internal kink mode that grows at re-collimation nozzles of collimated jets could lead to efficient magnetic energy dissipation, reducing the jet's magnetization parameter, $\sigma$, which is high before the flow enters the nozzle, down to $\sigma\approx 1$ \cite{bromberg2016}. At this point, the poloidal and toroidal magnetic field components in the frame co-moving with the jet are comparable. 

Kink instability has been studied both analytically \citep{rosenbluth1973,begelman1998,lyubarskii1999,appl2000,das2019} and using MHD simulations \citep{mizuno2009,mizuno2012,oneill2012,bromberg2019}. It triggers reconnection in current sheets, which dissipates magnetic energy into plasma energy. The importance of this process has been discussed in the context of GRBs \citep{drenkhahn2002,giannios2006,mckinney2012} and AGNs \citep{mckinney2009}. Kink instability has also been studied in laboratory experiments. For example, \cite{duck1997} observed a resonant kink mode, where $B_\phi/B_z \approx 1.0$ (hereafter, $z$ defines the direction along the jet's axis, and $\phi$ corresponds to a toroidal direction with respect to the same axis). Coincidentally, similar conditions are expected in collimation nozzles of relativistic jets. 

Particle acceleration in the process of kink instability was studied using PIC simulations by \cite{alves2018}. They considered a pressure supported jet where the toroidal magnetic field component dominates and found significant particle acceleration {\rm solely} due to the generation of an ideal coherent electric field along the jet axis. {\rm Since their setup is pressure supported,} force-balance implies,  $\nabla p = \vec{J}/c \times \vec{B}$, which effectively translates to $p \approx B^2/8\pi$ (hereafter, $p$ is the plasma pressure, and $\vec{J}=c\nabla\times \vec{B}/4\pi$ is the plasma current density). Therefore, their setup considers an effective, "hot", magnetization $\sigma_{\rm h} = {B^2}/{4\pi w} \approx 1$, where $w= \varepsilon+p$ is the gas relativistic enthalpy, and $\varepsilon$ is the plasma internal energy. AGN jets are, however, thought to be launched with  $\sigma_{\rm h}\gg1$ and exhibit force-free behavior close to their origin \citep{tchekhovskoy2016,bromberg2016}. Without an additional dissipation process, their cores will remain highly magnetized and cold until they become kink unstable at the jet nozzle. 

In \cite{bromberg2019} we studied the long-term evolution of the kink instability in force-free non-rotating jets using MHD simulations. We showed that the system relaxes to a Taylor state while conserving the net magnetic helicity and axial magnetic flux. Depending on the initial field configuration and the box size,  10--50\% of the magnetic energy is dissipated during the relaxation process. In this Letter, we investigate the mechanisms responsible for the particle acceleration during the process of kink instability by performing particle-in-cell (PIC) simulations. We consider the same magnetic field configurations as in \cite{bromberg2019} and study the regime of $\sigma_{\rm h} \gg 1$ and  $B_\phi/B_z \approx 1.0$. We find no coherent axial electric field in our setups, and find that particle acceleration occurs due to a combination of reconnection and turbulence.

\section{Numerical setup}
The first setup we consider is a force-free non-rotating jet originally investigated with MHD simulations by \cite{mizuno2009} and by \cite{bromberg2019}. The magnetic field profile consists of a strong vertical field, $B_{\rm z}$, dominated core surrounded by a region dominated by a toroidal field component, $B_\phi$. The magnetic field profile is given by,
 
 \begin{eqnarray}
 B_{\rm z} &=& \frac{B_0}{\left[1+(r/r_{\rm core})^2\right]^\zeta},\\
 B_{\phi} &=& B_{\rm z}\frac{r_{\rm core}}{r}\sqrt{\frac{\left[1+(r/r_{\rm core})^2\right]^{2\zeta}-1-2\zeta(r/r_{\rm core})^2}{2\zeta-1}},
 \end{eqnarray}
where $B_0$ is a scale factor that determines the value of magnetization parameter at the axis, $r_{\rm core}$ sets the size of the kink unstable core, and $r$ is the cylindrical radius. For $r\gg r_{\rm core}$ both field components asymptotically approach zero. The free parameter $\zeta$ sets the behavior of the magnetic pitch, $P=r B_z/ B_{\phi}$. For $\zeta<1$ the pitch is increasing with $r$, for $\zeta=1$ the pitch is constant, and for $\zeta>1$ the pitch is decreasing with $r$. In this work we consider two representative values of $\zeta$, $\zeta=0.64$ (Increasing Pitch, IP) and $\zeta=1.44$ (Decreasing Pitch, DP). The radial profile of the pitch is important for the global evolution of the instability. In the case where the pitch is increasing with the cylindrical radius, resonant surfaces confine the instability to the kink unstable core \citep{rosenbluth1973}, while in the case of a decreasing pitch profile the instability becomes disruptive.
 
We also consider a force-free setup by \cite{bodo2013}, which has a non-monotonic pitch profile and a strong confining vertical magnetic field outside of the kink unstable core. We term this profile as embedded pitch (EP, same as CO in \citealt{bromberg2019}). The magnetic field in this case is given by 
 \begin{eqnarray}
 B_\phi &=& \frac{B_0 R}{r}\sqrt{\left(1-e^{^{-(r/r_{\rm core})^4}}\right)},\\
 B_z &=& \frac{B_0RP_0}{r_{\rm core}^2}\sqrt{\left(1-\sqrt{\pi}\left(r_{\rm core}/P_0\right)^2{\rm erf}\left[(r/r_{\rm core})^2\right]\right)},
 \end{eqnarray}
where $R$ is the cylindrical radius of the domain's outer boundary, and the parameter $P_0$ is the value of the magnetic pitch at the axis. We consider a value of $P_0=1.5 ~r_{\rm core}$. The magnetic field configuration qualitatively differs from the IP and DP setups, since for $r>r_{\rm core}$ the axial component of the magnetic field, $B_z$, asymptotes to a constant value. This vertical magnetic field leads to a strong confinement of the jet. 

\begin{figure*}
  \centering  
    \includegraphics[width=0.90\textwidth]{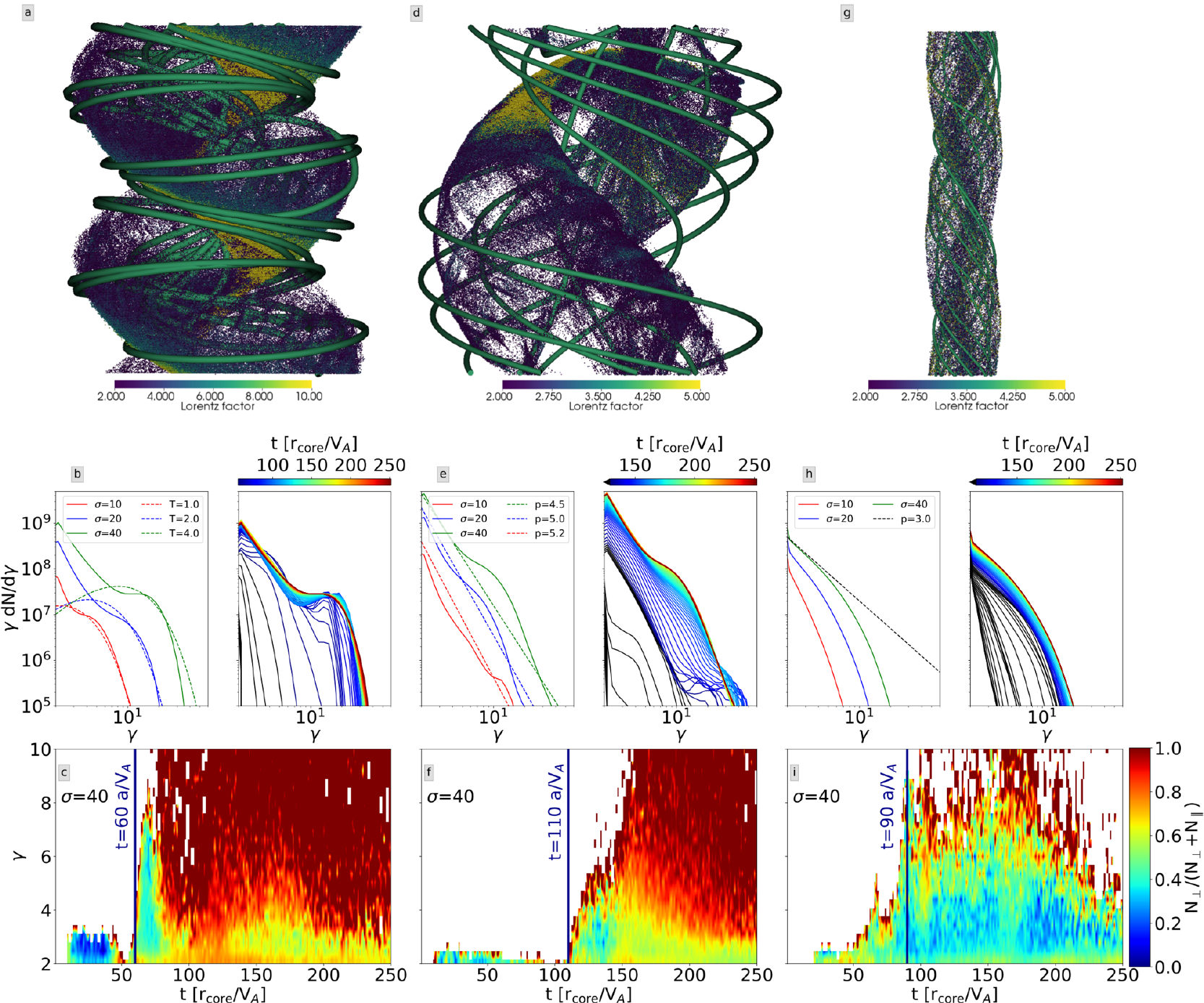}
      
    \caption{From left to right: decreasing pitch (DP), increasing pitch (IP), and embedded pitch (EP) cases. In the top row, thick green lines show magnetic field lines. Subsampled distribution of energetic particles is visualized as dots colorcoded by their Lorentz factors. Plots are computed at $t=60, 110, 90  ~r_{\rm core}/V_A$ correspondingly, the onset times of the acceleration episode in each configuration (see bottom panel). The middle row shows distribution functions (DFs) for all three setups, each set of two plots shows DFs at the end of the simulation on the left for all three $\sigma_0=10,20,40$ values, and the time evolution of the spectrum of the $\sigma_0=40$ run on the right. Panel b also includes Maxwellians fitted to the DFs, panel e, and h show power laws fitted to the DFs. The bottom row shows statistics of the acceleration events as a function of simulation time and particle energy. For a given particle at a particular energy, we classify the acceleration episode based on if parallel or perpendicular electric field dominates particle energization. $N_{\parallel}$ and $N_{\perp}$ are the numbers of parallel and perpendicular acceleration events, respectively. Initial particle distribution is a Maxwellian with a low temperature, $10^{-2}~m_e c^2/k_{\rm B}$, and all the spectra correspond to energized particles with $\gamma>2$.}
   \label{fig:3D}
\end{figure*}

We perform our simulations in the frame co-moving with the jet, thus the plasma is initially at rest. We use the relativistic PIC code {\tt Tristan-MP} \citep{tristancode}. The simulations are performed in a Cartesian three-dimensional computational box. The box length in $z$, $L_z$, is set to fit two wavelengths of the fastest-growing kink mode $\lambda_{\rm max}=8\pi P_0/3$ where $P_0$ is the value of the pitch at the axis \citep{appl2000}. We initialize our calculations with a cold uniform electron-positron plasma with temperature $T=10^{-2}~m_e c^2/k_{\rm B}$, and a density of ten particles per cell giving a total of $\sim10^{11}$ particles in the simulation box. We set both electrons and positrons to drift in opposite directions with velocities $\vec{v}_{\rm dr}=\pm \vec{J}/2 n e$ to generate the currents that support the initial magnetic field profile. The simulations are run up to $t=300 ~r_{\rm core}/V_{\rm A}$, where $V_A$ is the Alfv\'en speed defined as $V_{A}=c\sqrt{\sigma_0/(1+\sigma_0)}$, and $\sigma_0$ is the magnetization at the jet axis, $\sigma_0={B_0^2}/({4\pi m_{\rm e} n}c^2)$. We set $r_{\rm core}=60$ cells, and use grid sizes of: a) DP,  $3000^2 \times 900$, b) IP, $3000^2 \times 1300$ and c) EP, $1200^2 \times 1600$ in the $(x,y,z)$ directions respectively. We studied the dependence of our results on the scale separation by varying the ratio between the size of the kink unstable core and the plasma skin depth, $d_{\rm e}=c/\omega_{\rm p}$, where $\omega_p=\sqrt{{4\pi e^2 n}/{m_{\rm e}}}$ is the plasma frequency. We varied $d_{\rm e}$ from 3 to 6 cells. The simulations presented in this Letter use a scale separation of $r_{\rm core}/d_{\rm e}=20$, where $d_{\rm e}=3$ cells, which is sufficient to recover the overall MHD evolution (see \cite{bromberg2019} and Appendix A). In the z-direction, we apply periodic boundary conditions, while at the boundary in the x-y plane we have an absorbing layer for both fields and particles \citep{Cerutti2015}. For all three setups, we present simulations for three values of the magnetization parameter at the axis, $\sigma_0=10, 20, 40$, which correspond to $\beta=8\pi n T/B_0^2 =2(k_{\rm B}T/m_e c^2) (1/\sigma_0)=[20,10,5]\times 10^{-4}$. Larmor gyration period $2\pi/\sqrt{\sigma_0} \omega_p$ is resolved with at least a few time steps for all simulation setups.

\begin{figure*}
\centering
        \includegraphics[width=0.9\textwidth]{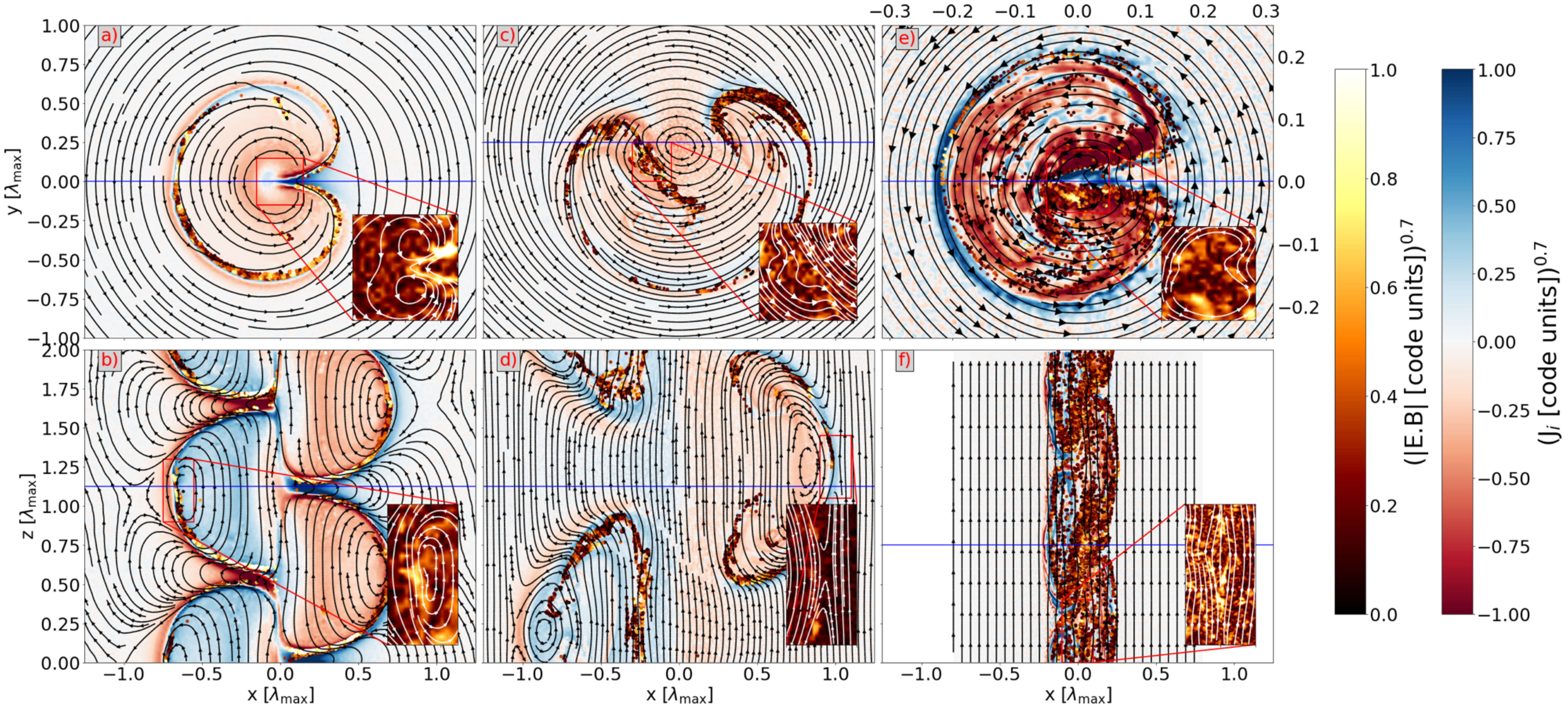}
   \caption{Formation of strong current layers in the onset of the non-linear stage of the kink instability. From left to right columns: DP, IP, and EP cases. First row: slices of the axial component of the current, $j_z$, in the x-y plane. Second row: slices of the toroidal component of the current, $j_{\phi}$, in the x-z plane. Black/white lines show the in-plane components of the magnetic field. Insets show the distribution of $E\cdot B$ as color and highlight the $E\cdot B \neq 0$ regions where in-plane magnetic field components show anti-parallel orientation. A subsample of particles with $\gamma>2$ is shown as dots, colorcoded with the local $E\cdot B$ values they experience. Their locations clearly correlate with strong current layers. The $E\cdot B$ colorbar is assigned to both the insets and the particle colorcoding.}
    \label{fig:currents}
\end{figure*}

\begin{figure}
    \centering
    \includegraphics[width=0.45\textwidth]{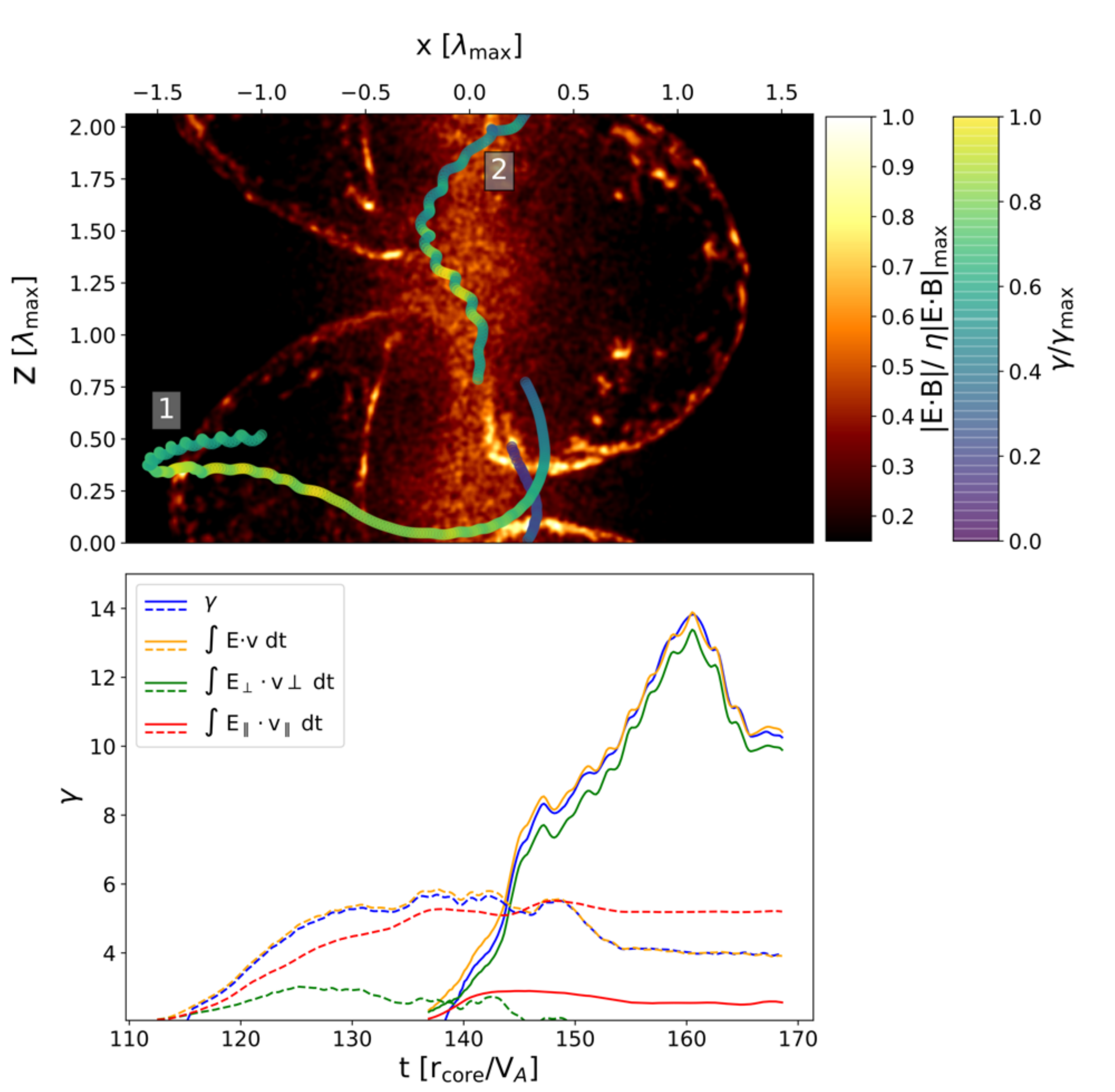}
    \caption{Trajectories of two accelerated particles in the IP case. Top panel shows $E\cdot B$ in the x-z plane, overplotted with trajectories of a particle (1) that undergoes mainly parallel acceleration, and a particle (2) that undergoes perpendicular acceleration. Lower panel shows the time-integrated work of the electric field, $E\cdot v$, along the trajectory of these particles, the contribution of parallel and perpendicular components to the integrated $E\cdot v$, and particle Lorentz factors as a function of time. The dashed lines correspond to particle $1$, and solid lines correspond to particle $2$. Particle $1$ is predominantly accelerated by a parallel electric field in the current layer at the edge of the kink unstable core, while particle $2$ experiences strong acceleration by perpendicular electric fields in the jet's core.}
    \label{fig:trajectories}
\end{figure}

\section{Results}
Our PIC simulations show the same global behavior found in our MHD simulations \citep{bromberg2019}. The sufficiently large separation between fluid and kinetic scales allows us to obtain similar growth-rates in the linear stage, and a comparable amount of electromagnetic energy dissipation as in the MHD simulations (between 15-20\% of the initial electromagnetic energy in all three setups, see appendix A). Initially, the most unstable mode is a kink mode with a longitudinal wavenumber $l=2$, and an azimuthal wave number $m=-1$\footnote{The longitudinal wavenumber is defined as $l=k_z L_z/2\pi$, where $k_z$ is the component of the wave vector in the $z$ direction. The azimuthal wavenumber, $m$, defines the type of mode, where modes with $|m|=1$ are known as kink modes. For a more detailed discussion on the properties and behavior of the unstable mode, see \citealt{bromberg2019}}. It gives rise to a global helical current sheet at the edge of the kink unstable core. Later on the $l=2$ mode transforms into an $l=1$ mode. Eventually, the global current sheet breaks up generating small-scale current sheets and turbulence that mediate further dissipation of the magnetic energy. A similar behavior was observed in our MHD simulations. 

In all three setups, we observe particle energization due to an electric field that is parallel (non-ideal) or perpendicular (ideal) to the local magnetic field direction. As the instability becomes non-linear, we observe a strong burst of particle energization due to a non-ideal electric field, which takes place in current sheets at the jet's periphery (see Fig. \ref{fig:3D} for a 3D visualization of a down-sampled distribution of simulation particles color-coded by their Lorentz factors). These sheets have strong guide fields, which are comparable to the reconnecting field at the periphery and can become up to $\sim 5$ stronger than it at the core. The presence of a strong guide field suppresses particle acceleration and leads to the formation of steep power laws in the particle distribution function (hereafter, DF). \cite{werner2017} studied relativistic reconnection in pair plasmas with strong guide fields using local PIC simulations, and found a relation between the strength of the guide field and the power law index, ${\alpha}$, of the DF, $f(\gamma)\propto \gamma^{-\alpha}$. In our work we find $\alpha \approx 3-5$, which is in agreement with their results for comparable strengths of the reconnecting and guide magnetic field components. At this stage, we find the maximum energy of accelerated particles to scale as $\gamma_{\rm max} \approx \chi r_{\rm core}/r_{\rm L0}$, where $r_{\rm L0}=m_e c^2/eB_0$ is a nominal cold relativistic gyroradius, and $\chi\approx 1/6$ \footnote{This conclusion is based on our simulations with different strengths of the jet's magnetic field. Increasing the jet's size is numerically expensive in our current setups, as the jet significantly expands laterally during the simulation time. We will conduct a systematic study of the dependence of $\gamma_{\rm max}$ on the jet's size in the future work.}. 

In all our setups, we find that the self-excited turbulence is small-amplitude, e.g. the mean field is stronger compared to the fluctuating component. We evaluate the amplitude of turbulence as $\xi=|(B(\vec{x})-\langle B(\vec{x}) \rangle|/\langle B(\vec{x})\rangle$, where $\langle B(\vec{x}) \rangle=\int B(\vec{x}') e^{- \mid \vec{x} - \vec{x}'\mid^2/2\sigma_{\rm std}^2} ~{\rm d}\vec{x}'$ is the magnetic field strength averaged with a Gaussian kernel, and $\sigma_{\rm std} = r_{\rm core}/3$.  We varied the size of the kernel in the range $\sigma_{\rm std} \in [r_{\rm core}/6 , r_{\rm core}/2]$ and found no qualitative differences in our conclusion based on this analysis. The value of $\xi$ itself varies spatially. We quantify the amplitude of turbulent motions by measuring the range of $\xi$ inside the kink unstable core. In all three setups we find $\xi \leq 0.1$. The small-amplitude turbulence leads to heating of the plasma, which forms a secondary Maxwellian in the DF (see panels b and e in Fig.~\ref{fig:3D}). The temperature of this Maxwellian scales with the initial magnetization parameter, namely, $k_{\rm B}T/m_e c^2\propto \sigma_0$. Particle energization at this stage is dominated by the perpendicular component of the electric field. To quantify the importance of both parallel and perpendicular electric fields during the evolution of the instability we trace every tenth particle in our simulations with $\gamma > 2$. We classify individual acceleration events based on if the parallel or the perpendicular electric field component dominates the acceleration by looking at the absolute values of energy gained by each process. The statistics of acceleration episodes are shown in Fig. \ref{fig:3D}, bottom row. In all three setups, a large fraction of the particles undergo parallel acceleration immediately after the instability becomes non-linear, while in the IP and DP case the perpendicular acceleration dominates at larger energies. We find that the number of acceleration events due to the parallel electric field increases at higher values of the magnetization parameter.

In Fig.~\ref{fig:trajectories} we show an example of two particle trajectories in the IP case that exhibit acceleration due to either a parallel or a perpendicular electric field. In the case of parallel acceleration (particle 1), the energization happens in the current sheet at the edge of the kink unstable core, where $E\cdot B \neq 0$. In the perpendicular case (particle 2), the particle is initially accelerated by a parallel electric field and then ends up in the turbulent core, where it undergoes further acceleration to higher energies mediated by the perpendicular electric field. These particle trajectories are representative for all three setups, although the relative contribution of parallel and perpendicular episodes differs, as can be seen in Fig. \ref{fig:3D}.

The DP simulation shows a strong acceleration event around $t=60 ~r_{\rm core}/V_A$, as is shown in Fig. \ref{fig:3D}c. At this time the $l=2$ mode forms a helical current sheet at the edge of the kink unstable core, see Fig. \ref{fig:3D}a. The sheet is produced by the relative shear between the magnetic field inside the jet's core and at the periphery and is supported by strong currents (see Fig. \ref{fig:currents}a). These current layers contain most of the energized particles and correlate with locations where $E\cdot B \neq 0$. In these layers, some of the magnetic field components exhibit anti-parallel orientations, see inset in Fig. \ref{fig:currents}d where $B_{\rm z}$ is the reconnecting field component. This shows that non-ideal electric fields in current sheets are the driving mechanism of the energization. The statistics of acceleration events in the DP case is shown in Fig. \ref{fig:3D}c, where the burst of acceleration events at $t=60 ~r_{\rm core}/V_A$ coincides with the increasing number of non-thermal particles in the DF (see Fig. 1b, right panel). Clearly, a majority of the particles is initially accelerated via parallel electric fields. At later times a second acceleration stage due to a perpendicular electric field in turbulence pushes the particles to higher $\gamma$ values. For all three values of $\sigma_0$, the DF shows the growth of a secondary Maxwellian with a temperature that scales linearly with $\sigma_0$, as is expected from the energy conservation argument\footnote{The Larmor radius of particles with $\gamma=\sigma_0$ in the jet's core is $r_{\rm L}=\sigma_0 r_{\rm L0}=\sqrt{\sigma_0}d_e$, which corresponds to $0.3 r_{\rm core}$ for $\sigma_0=40$. The size of the kink unstable core, however, grows to $\sim 0.75 \lambda_{\max}\sim 10 r_{\rm core}$ in the non-linear stage, so further particle acceleration is in principle possible. The plasma skin depth, $d_e=\sqrt{m_e c^2 \langle \gamma \rangle /4\pi e^2 n}$, also increases as a result of the heating (see also Appendix A for the discussion of the scale separation in the DP case).}. The measured amplitude of the turbulence for $\sigma_0=40$ is of the order of $\xi\leq 0.1$ in the kink unstable core. 

In the IP case, the first acceleration event is seen at $t=110 ~r_{\rm core}/V_A$. At this time, the $l=1$ mode develops a current sheet at the jet's periphery. Again, the location of particle acceleration correlates with current sheets where $E\cdot B \neq 0$, as can be seen in Fig. \ref{fig:currents}. The statistics of acceleration events in Fig. \ref{fig:3D}f clearly shows that at this time, the majority of particles are accelerated due to parallel electric fields. The resulting spectra in Fig. \ref{fig:3D} shows a power law with $\alpha\approx 4.5$ for $\sigma_0=40$, and a secondary Maxwellian that slowly grows over time. We measure the amplitude of turbulence in the core to be of the order of $\xi \leq 0.05$, which is smaller compared to the DP simulation. This can explain the slower growth of the secondary Maxwellian in the spectra.

For the EP case, at $t=50 ~r_{\rm core}/V_A$ the particle acceleration starts when the $l=2$ mode grows. Again, current sheets coincide with locations of $E\cdot B \neq 0$, where particles are accelerated due to parallel electric fields. The resulting DF shows a clear power-law with index $\alpha \approx 3$ for $\sigma_0=40$, and a modest steepening of the spectrum for lower values of $\sigma_0$. The turbulence in the EP setup has a small-amplitude, of the order of $\xi\leq0.01$, which could explain the lack of a secondary Maxwellian in the spectra. This correlates with a strong dominance of parallel acceleration events in the particle energization history, which takes place over the entire simulation duration in the EP case, as shown in Fig. \ref{fig:3D}i.

\begin{figure*}
    \includegraphics[width=0.95\textwidth]{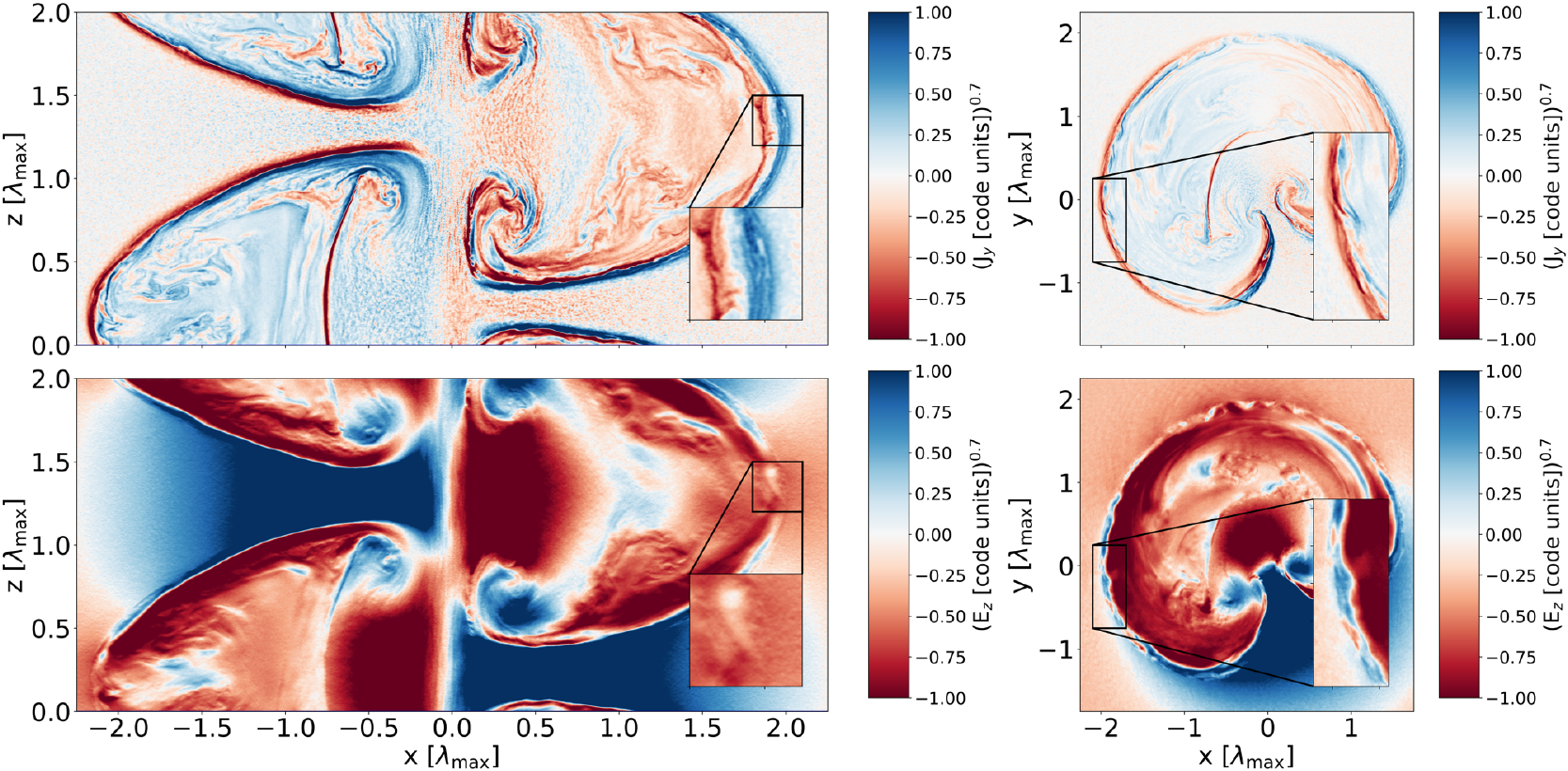}
    \caption{Formation of plasmoids in the IP setup. The first row presents the $y$ component of the current in x-z and x-y planes. The second row shows the $z$ component of the electric field. In all panels, insets zoom into plasmoid-like structures. In all panels distances are measured in units of the fastest-growing kink mode $\lambda_{\rm max}=8\pi P_0/3$, where  $P_0$ is the value of the pitch on the axis.}
   \label{fig:plasmoids}
\end{figure*}

Thin current sheets are known to be unstable to a tearing instability, and subsequent plasmoid instability of secondary sheets \citep{loureiro2007}. While limited scale separation of our global simulations prevents us from observing the plasmoid instability, we do observe the initial tearing of current sheets generated by the relative shear of the magnetic field at the jet's boundary. An example of the IP case is presented in Fig. \ref{fig:plasmoids}, where different quantities show plasmoid-like structures in different parts of the current sheet at the jet's boundary. We plan to study kink-unstable configurations presented in this work with relativistic resistive MHD simulations with adaptive mesh refinement \citep{Ripperda2017}, in order to better resolve plasmoid chains in these current sheets. 

\section{Discussion and conclusion}

 Reconnection and turbulence in collisionless plasma were studied so far in idealized periodic boxes. Our study shows how they can be self-consistently excited and energize particles in the process of kink instability in highly magnetized jets. We find that acceleration in current sheets dominates at low particle energies; and happens due to non-ideal electric fields that lead to the formation of steep power laws in the DF, due to strong guide fields at the reconnection sites. The presence of acceleration due to non-ideal electric fields is in contrast with the study of \cite{alves2018}. This difference is likely caused by the fact that their pressure-supported jet configuration corresponds to the case $\sigma_{\rm h} \approx 1$. As we discuss above, we also find no coherent axial electric field in our highly magnetized, force-free setups. 
 
 While we observe plasmoid formation, our limited scale separation does not allow the formation of a full plasmoid chain, and a study of the Fermi-like process of particle acceleration in plasmoids \citep{petropoulou2018}. Future large-scale local simulations of reconnection with a strong guide field are needed to investigate this potentially important mechanism of particle acceleration \citep{Drake06}. We further find that energization due to scatterings on small-amplitude turbulent fluctuations leads mostly to plasma heating. This is in contrast to local simulations of particle energization in high-amplitude turbulence \citep{zhdankin2013,zhdankin2017,comisso2018}, which showed formation of prominent power laws. Motivated by our results in the DP case, where particle energization in turbulence erases the initial reconnection spectra, for the cases of large-amplitude turbulence we anticipate the power laws to extend up to energies corresponding to the confinement condition, $\gamma_{\max}\sim r_{\rm core}/r_{\rm L0}$ \citep{zhdankin2017}. 
 
 Future work should incorporate realistic jet structures, including rotation and velocity shear, and develop an understanding of how to extrapolate the results of simulations with limited scale separation, such as ours, to parameters of astrophysical systems. Similarly to this work, these studies will identify the geometry of current sheets and quantify the amplitude of the excited turbulence and thus, allow to quantify particle acceleration and emission of energetic photons from kink-unstable jets in GRB and AGN from {\rm first principles}. 

\acknowledgments

The authors thank A. Bhattacharjee, L. Comisso, H. Hakobyan, B. Ripperda, L. Sironi, A. Spitkovsky and A. Tchekhovskoy  for  insightful comments over the course of this project. J.D. is funded by the ERC Synergy Grant 610058, \cite{goddi2017}. The authors thank the anonymous referee for insightful comments. O.B. and C.S. were funded by an ISF grant 1657/18 and by an ISF (I-CORE) grant 1829/12. O.B. and S.P. were also supported by a BSF grant 2018312. S.P. acknowledges support by the National Science Foundation under Grant No. AST-1910248. The Flatiron Institute is supported by the Simons Foundation. 
\newpage
\software{ {\tt Tristan-MP} \citep{tristancode}, {\tt python} \citep{travis2007,jarrod2011}, {\tt scipy} \citep{jones2001}, {\tt numpy} \citep{walt2011}, {\tt matplotlib} \citep{hunter2007}, {\tt VisIt} \citep{visit}.}

\appendix
\section{Comparison with MHD}

In order to ensure that our simulations probe the large-scale behavior correctly, we compare the growth rates of the kink instability and electromagnetic dissipation rates of our PIC simulation with MHD simulations of the same configurations from \cite{bromberg2019}. The simulation box sizes are identical, and we choose $\sigma_0=10$, the separation between the size of the kink-unstable core and the plasma skin depth in the case of PIC $r_{\rm core}/d_e$=20, for this comparison. To compute dissipation rates in both PIC and MHD simulations, we correct for the electromagnetic energy that leaves through the box boundary $\mathcal{A}$ (edge of the absorbing boundary for PIC, and the edge of the box with standard outflow boundary condition in the case of MHD). 

\begin{figure*}
  \centering
    \includegraphics[width=0.85\textwidth]{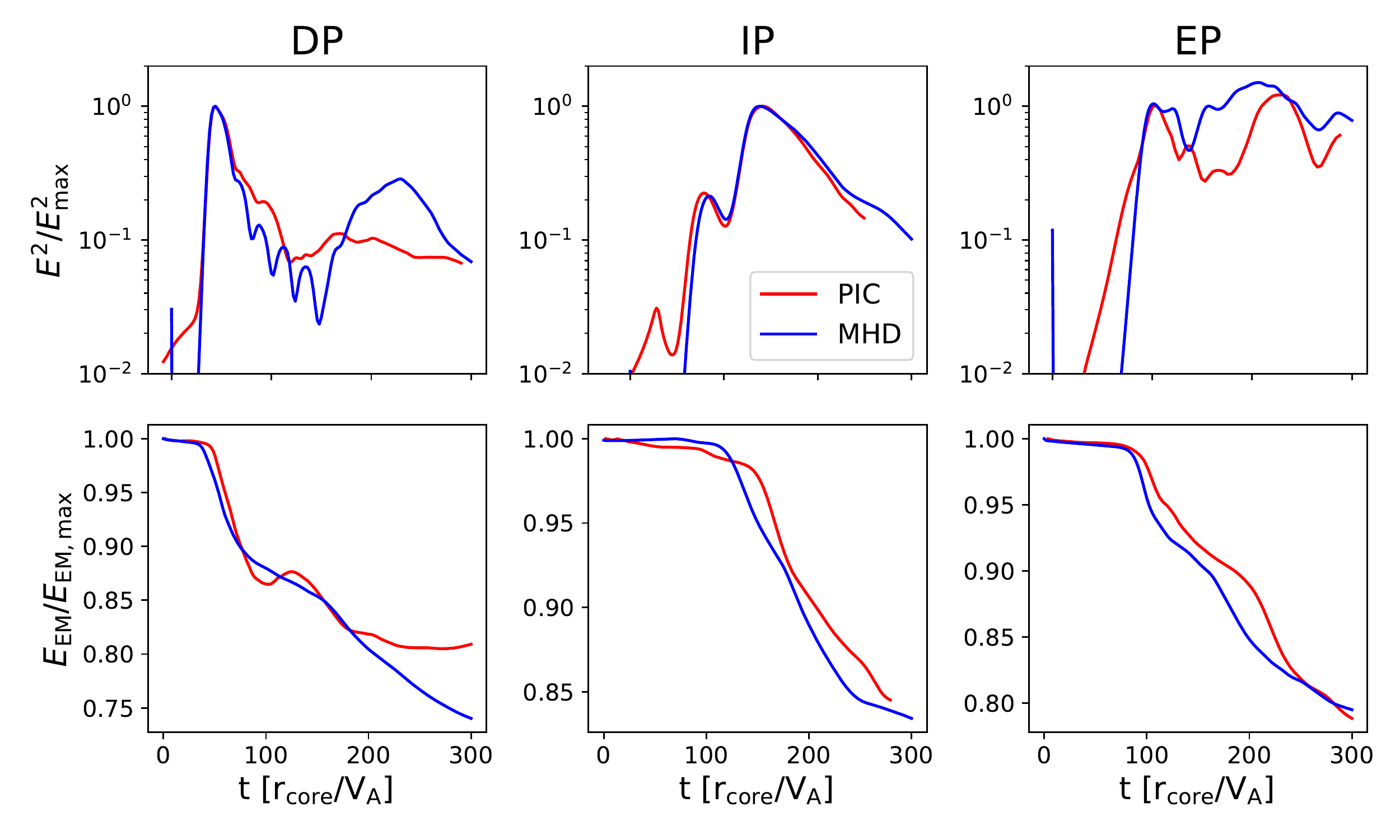}
    \caption{Comparison of the linear growth rates of the instability and electromagnetic energy dissipation in PIC and MHD simulations. From left to right: DP, IP, and EP case. Panels in the top row show the evolution of electric energy as a function of time, which highlights a stage of exponential growth. Panels in the bottom row show the dissipation of electromagnetic energy. In all panels, red lines represent PIC simulations, and blue lines correspond to MHD simulations.}
   \label{fig:comp}
\end{figure*}

The growth rates of the electric energy are shown in the top panels of Fig. \ref{fig:comp}. In the PIC simulations, the onset of the instability is slightly delayed with respect to MHD. We, therefore, shifted the PIC curves so that they overlap with the MHD curves to ease the comparison of the rates by eye. The linear growth shows very similar rates in PIC and MHD. In the PIC simulations, the instability initially kicks in on kinetic scales at the jet's boundary, which is not observed in the MHD simulations. This behavior is significantly more prominent in simulations with $r_{\rm{core}}/d_e=10$, which highlights the importance of using large scale separation in PIC simulations. The small scale plasma instabilities cause some discrepancies between the linear growth rates at the very early times. Also, the initial amplitude of the electric field is higher in the PIC runs because of the particle noise. However, when the kink instability grows and the jet expands at $t\geq 50 r_{\rm{core}}/V_A$, the growth rate in PIC becomes indistinguishable from the one observed in MHD (see \cite{bromberg2019} for MHD simulations). At this stage, the growth rates are observed to be nearly identical in PIC and MHD for all three setups. The magnetic field dissipation is shown in the bottom row of Fig. \ref{fig:comp}. In the DP, IP, and EP cases, the evolution and dissipation rates up to $t=200r_{\rm core}/V_{A}$ are very similar. This comparison shows excellent agreement between the large scale behavior of the kink instability in the PIC simulations presented here and the MHD simulations from \cite{bromberg2019}. In the DP case the MHD simulation continues to dissipate, while PIC saturates at around $E_{\rm{EM}}/E_{\rm{EM,t=0}}/\approx 0.8$. The discrepancy is likely due to the fact that the separation between the jet scale and the skin depth scale shrinks because of the plasma heating during the turbulent stage of the instability, which is most prominent in the DP case. For our box size, $L_z=2\lambda_{\max}$, running MHD simulations further does not lead to the larger amount of the dissipation for the cases of IP and EP. However, the final state is not fully relaxed, and at least twice larger simulation box (in all directions, since the jet also expands more prominently if a larger amount of most unstable modes is present) is required to observe a Taylor state \citep{bromberg2019}. About twice the amount of dissipation is observed in simulations, which lead to full relaxation, $\sim 30\%$ and $\sim 40\%$ for the IP and EP cases, correspondingly (see Fig. 6 in \citealt{bromberg2019}). A significantly higher, up to $40\%$, amount of dissipation is observed in MHD simulations of the DP case with the same box size as chosen in this Letter. As we mentioned above, the decreased scale separation is a likely reason for this discrepancy.

\end{document}